\documentclass{article}[12pt]
\usepackage[bookmarks=false]{hyperref}
\usepackage{graphicx}
\usepackage{amssymb}
\begin{document}

\begin{center}
{\noindent\Large
Phase transitions and rare-earth magnetism in hexagonal and
orthorhombic DyMnO$_{3}$ single crystals
}
\end{center}
\vspace{0.4cm}
S Harikrishnan$^{1}$, S R\"o\ss{}ler$^{2}$,
C M Naveen Kumar$^{1}$, H L Bhat$^{1,4}$, U K R\"o\ss{}ler$^{3}$, S
Wirth$^{2}$, F Steglich$^{2}$, and Suja Elizabeth$^{1}$
\\
$^{1}$Department of Physics, Indian Institute
of Science, Bangalore-560012, India \\
$^{2}$ Max Planck Institute for Chemical Physics of
Solids, N\"othnitzer Stra\ss{}e 40, 01187 Dresden, Germany \\
$^{3}$ IFW Dresden, Postfach 270016, D-01171 Dresden,
Germany \\
$^{4}$ Centre for Liquid Crystal Research, Jalahalli, Bangalore-560013, India
\\
%
\begin{abstract}
%
The floating-zone method with different growth ambiances
has been used to selectively obtain hexagonal or orthorhombic
DyMnO$_{3}$ single crystals.
The crystals were characterized by X-ray powder diffraction of
ground specimen and a structure refinement as well as electron
diffraction.
We report magnetic susceptibility, magnetisation, and
specific heat studies of this multiferroic compound
in both the hexagonal and the orthorhombic
structure.
The hexagonal DyMnO$_{3}$ shows magnetic
ordering of Mn$^{3+}$ (S =2) spins on a triangular
Mn lattice at $T^{\mathrm {Mn}}_{N}$ = 57~K characterized by a
cusp in the specific heat.
This transition is not
apparent in the magnetic susceptibility
due to the frustration on the Mn triangular lattice
and the dominating paramagnetic susceptibility
of the Dy$^{3+}$ (S=9/2) spins.
At $T^{\mathrm {Dy}}_{N}$ = 3~K, a partial antiferromagnetic order
of Dy moments has been observed.
In comparison, the magnetic data for orthorhombic DyMnO$_{3}$
display three transitions.
The data broadly agree with results from earlier neutron
diffraction experiments, which allows for the following
assignment: a transition from an incommensurate antiferromagnetic
ordering of Mn$^{3+}$ spins at $T^{\mathrm {Mn}}_N$ = 39~K, a
{\textit{lock--in}} transition at $T_{\mathrm {lock-in}}$ = 16~K
and a second antiferromagnetic transition at $T^{\mathrm {Dy}}_N$
= 5~K due to the ordering of Dy moments.
Both the hexagonal and the orthorhombic crystals show magnetic
anisotropy and complex magnetic properties
due to 4$f$-4$f$ and 4$f$-3$d$ couplings.\\
\end{abstract}

\newpage
\section
{
Introduction
}
%
Multiferroic manganites of the type $R$MnO$_3$ ($R$=rare-earth) 
have attracted a great deal of attention due to two main reasons:
First, on the technological front multiferroics are promising
materials for potential spintronic applications such as four-state
memory devices \cite{Bibes08}.
For their application, however, a better understanding of their
fundamental properties is necessary.
Second, these materials belong to a class of
complex oxides with various coupled ordering phenomena
and rich phase diagrams
\cite{Gajeknature07,Eerensteinnature06,Prellierjpcm05}.
Generally, $R$MnO$_3$ crystallizes in an orthorhombic,
perovskite-like structure (space group $Pnma$), if the rare-earth
ionic radius is large enough, i.e.\ $r_R > r_{\mathrm {Dy}}$
\cite{Zhouprb06}.
Thus, the manganites with $R$ = La, Pr, Nd etc.\ are orthorhombic,
whereas the $R$MnO$_3$ manganites for $R$ = Y, Tm, Yb, Lu, Er
with an ionic radius smaller than that of Dy usually crystallize in hexagonal
structure (space group $P6_{3}cm$).
Since the transition between hexagonal and orthorhombic crystal
structure of $R$MnO$_3$ occurs in the rare-earth series for the
ionic radius of Dy or Ho, $R$MnO$_3$ with $r_R \simeq  r_{\mathrm
{Dy}}$ can be synthesized in both structures by applying different
growth conditions \cite{Zhouprb06}.

Multiferroic $R$MnO$_3$ compounds are simultaneously ferroelectric
and magnetic.
Both hexagonal as well as orthorhombic $R$MnO$_3$ exhibit
multiferroic properties but the origin of the spontaneous
polarisation is different in these two symmetries.
In the hexagonal system, the structural distortion arising from an
asymmetric coordination of oxygen around $R$ leads to a net
polarisation, as exemplified in the case of YMnO$_3$
\cite{Leeprb05}, and the ferroelectric transition temperatures are
as high as $T_{FE} > $~700~K.
Hexagonal \mbox{({\it h-})$R$MnO$_3$} exhibits two kinds of
magnetic phase transitions: (i) an antiferromagnetic ordering of
the Mn-sublattice below a  N\'{e}el temperature
$T_N~{\sim}$~100~K, and (ii) a low temperature ordering of the $R$
moments usually below 10~K \cite{Tomuta03}.
The ferroelectric order persists down to low temperatures through
these magnetic transitions \cite{Lonkaiprb04}, and a coupling
between antiferromagnetic and ferroelectric domains has been
observed by optical second harmonic generation (SHG)
\cite{Fiebigprl02}.
The crystal structure of {\it h-}$R$MnO$_3$ essentially consists
of corner-sharing MnO$_5$ bipyramids separated by layers of
$R^{3+}$.
The Mn ions within the $ab$ plane of the hexagonal structure
occupy a stacked triangular lattice giving rise to an
antiferromagnetic Mn-O-Mn (super-)exchange coupling between
nearest Mn-neighbours \cite{Bertautpl63,Munozcm01}.
Below 100~K a three-dimensional ordering of the Mn spins results in an antiferromagnetic phase in {\it h-}$R$MnO$_3$
\cite{Tomutajpcm01}.
This antiferromagnetic transition has been observed in magnetic
susceptibility and specific heat of {\it h-}$R$MnO$_3$ systems
like ScMnO$_3$, LuMnO$_3$ and YMnO$_3$ \cite{Tomutajpcm01}.
At the  N\'{e}el temperature, a non-collinear 120$^\circ$
antiferromagnetic spin structure is established owing to the
geometrical frustration of the triangular lattice planes.
The coupling of these triangular lattices of Mn-layers along the
$c$ axis is difficult to resolve from macroscopic measurements but
it has been investigated for various hexagonal manganites, e.g.,
by neutron diffraction and SHG methods \cite{Fiebigprl00}.
The responsible and much weaker Mn-O-O-Mn super-superexchange
couplings may be ferromagnetic or antiferromagnetic depending on
the particular rare-earth through changes of the lattice
parameters \cite{Lonkaijap03}.
In the much-studied compound {\it h-}HoMnO$_3$ from this class of
materials, an antiferromagnetic coupling has been proposed from
neutron diffraction \cite{Munozcm01}.

Generally, the $4f$ moments of $R^{3+}$ in {\it h-}$R$MnO$_3$
may order at very low temperature.
The interaction of the magnetic sublattices of {\it
h-}$R$MnO$_{3}$ poses interesting problems due to the frustration
in the Mn sublattice which carries over to the coupling of
$R$-moments within their $R$ sublattice \cite{Fiebigprl02}.
It should be noted that the rare-earth ions occupy two
inequivalent crystallographic positions -- the 2(a) and 4(b) sites
denoted as $R1$ and $R2$, respectively -- while the Mn ions occupy
the 6(c) position.
Hence, the rare-earth magnetism in {\it h-}$R$MnO$_3$ may display
multi-sublattice effects, and the ordering of the rare-earth
moments in the two different crystallographic sites 4(b) and 2(a)
is found to be different for different rare-earths.
Moreover, the magnetic moments on these two sites can be
considered to be independent and do not have to be of the same
size \cite{Tomuta03, Munozcm01}.
For different rare-earth ion, the geometric arrangement of the
sublattice remains the same, but the low temperature magnetic
properties can be different depending on the anisotropy
as well as the sign and relative
strength of the 4$f$-4$f$ and 4$f$-3$d$ couplings.
However, the rare-earth moments at the 4(b) sites are usually
ordered and directed perpendicular to the $ab$ basal plane, i.e.,
parallel to the $c$-axis. Hence, the inter sublattice coupling may
be frustrated too and may give rise to complex magnetic phase
diagrams.

In {\it h-}HoMnO$_{3}$ the Mn spins order at $T^{\mathrm {Mn}}_N$
$\approx$ 72~K \cite{Vajkprl05} and, at temperatures below this
transition, this compound displays a complex magnetic phase
diagram (as studied recently by Lorenz {\it et~al.}
\cite{Lorenzprb05}).
The lattice anomalies introduced by the magnetoelastic
interactions of the novel magnetic phases in HoMnO$_3$ are also
reflected in dielectric measurements as anomalies \cite{Yenprb05}.
The fact that the magnetic structure of this class of $R$MnO$_3$
is complex and poorly understood is reinstated by the ongoing
attempts to determine the magnetic structure and the magnetic
phase diagram including an intricate succession of field-induced
phases in HoMnO$_3$ \cite{Brownprb08,Nandiprl08}.
In comparison, relatively little is known about the magnetism of
{\it h-}DyMnO$_3$ \cite{Ivanov06}.
Therefore, a detailed investigation on hexagonal DyMnO$_3$ may
provide further insight into the magnetism of this class of
materials.

In orthorhombic ({\it o-})$R$MnO$_3$ multiferroics with
perovskite-like crystal structure the ferroelectricity is
attributed to a complex spiral spin-order that breaks the
inversion symmetry \cite{Cheongnature07,Kimuranature03}.
The associated magnetically driven transitions usually occur at much lower
temperatures.
{\it o-}DyMnO$_3$ has been investigated in some detail mainly by
diffraction methods, and hence, detailed microscopic information
on the spin-structure is available
\cite{Feyerhermprb06,Kimuraprb05,Prokhnenkoprl07,Gotoprl04}.
An incommensurate ordering of Mn moments sets in at $T^{\mathrm
{Mn}}_N = $~39~K.
The collinear magnetic order is of a longitudinal sinusoidal
spin-density-wave type with moments along the $b$-axis in the
$Pbnm$ space group setting and a  propagation vector
$(0,q_{\mathrm {Mn}},0)$ that varies with temperature below
$T^{\mathrm {Mn}}_N$ \cite{Feyerhermprb06,Kimuraprb05}.
At $T_{\mathrm {lock-in}}=$~18~K, an additional component of the
Mn moment along the $c$ axis gives rise to a spiral (cycloidal)
magnetic order and breaks the inversion symmetry
\cite{Prokhnenkoprl07, Gotoprl04}.
At this temperature, spontaneous electric polarization is
observed along the $c$ axis.
This electric polarization undergoes a flop transition from the
$c$ to the $a$ axis when a magnetic field is applied within the
$ab$ plane.
Below $T^{\mathrm {Dy}}_N$ = 5~K, the Dy moments order in a
commensurate structure with propagation vector along $b$ \cite
{Feyerhermprb06}.
Strong structural distortions related to the ordering of Mn and Dy
magnetic moments and couplings between these moments have been
observed by synchrotron x-ray diffraction (XRD) and resonant
magnetic scattering experiments \cite{Feyerhermprb06}.

Here, we report magnetic susceptibility, magnetisation and
specific heat measurements on both {\it h-}DyMnO$_3$ and {\it
o-}DyMnO$_3$ single crystals.
Different crystal growth atmospheres have been used to stabilize
the system in one of these two crystal structures.
The macroscopic measurements of magnetic properties and specific
heat broadly agree with the microscopic information on the spin
structure and magnetic phase transitions in the  DyMnO$_{3}$
system.
The results in particular show that the Dy magnetic moments at low
temperatures are only partially ordered in the case of {\it
h-}DyMnO$_{3}$, whereas they are completely ordered
in the case of the {\it o-}DyMnO$_{3}$.\\
%
\section
{
Experimental
}

%
Single crystals of DyMnO$_3$ were grown by the optical
floating-zone method in an infrared furnace (FZ-T-10000-H-VI-VP
procured from Crystal Systems Inc.).
The furnace is equipped with four hemi-ellipsoidal mirrors and
halogen lamps that are capable of delivering a total power of 6
kW.
The starting materials for growth were prepared following the
standard route of solid state reaction.
The precursors Dy$_2$O$_3$ and MnO$_2$ were mixed intimately and
subsequently heat treated at 1200$^0$C for 12~h.
Then, the material was reground and annealed again at 1250$^0$C
for 24~h.
This process was repeated until a single phase was obtained.
X-ray diffraction (XRD) was performed to confirm phase purity.
The powder was used to fabricate ceramic ingots for crystal growth
by filling it into rubber tubes and subjecting them to a
hydrostatic pressure of 70~MPa.
These cylindrical ingots were sintered at 1450$^0$C for 12~h prior
to growth.
During the growth the seed and feed rods were rotated at 40~rpm in
opposite directions.
At a growth rate of 4 - 6~mm/h crystals of typical dimensions 5~cm
length and 4~mm - 6~mm diameter were obtained.
Growth was performed in the ambiance of either argon or air, to
obtain either hexagonal or orthorhombic crystals, respectively.
It should be noted that the Gibb's free energy of the two
polymorphic modifications of DyMnO$_{3}$ are close
\cite{Ivanov06}
and though the perovskite phase is stable at room temperature, a transformation to
a hexagonal phase is possible at temperatures greater than 1600$^{\circ}$C
\cite{Szabo69}. 
It is documented that the synthesis in oxygen leads to the formation of a perovskite phase
whereas oxygen deficit leads to the hexagonal phase
\cite{Alonsoic00}.
In an oxygen-deficient atmosphere the crystal will be unable to complete the 
twelve-fold coordination of $R$ required for the perovskite; instead an eight-fold coordination
with hexagonal symmetry results. 
Laue photographs of the grown crystals indicated their single
crystalline nature.
Chemical composition of these crystals was determined by energy
dispersive X-ray analysis (EDAX) as well as by inductively coupled
atomic emission spectroscopy (ICPAES) using a Perkin Elmer
Spectrometer Optima 2000.
Powder X-ray diffractograms of pulverized samples were obtained
from a Philips X'Pert diffractometer with Cu-K${\alpha}$ radiation
(${\lambda}$ = 1.54 {\AA}).
Slow scans with a resolution of 0.01$^{\circ}$ were obtained in
the range 2${\theta}$ = 5--100$^{\circ}$.
Crystal structure refinement was performed by the Rietveld method
\cite{Rietveld69} using the FULLPROF code \cite{Carvajal}.
Selected area electron diffraction (SAED) patterns were obtained
through transmission electron microscopy using a Tecnai G 30
electron microscope.
Magnetic measurements were conducted on oriented single crystals
in a commercial (Quantum Design) superconducting quantum
interference device magnetometer (SQUID) in the temperature range
2~K -- 300~K.
Magnetic susceptibility and specific heat measurements were performed
using a physical property measurement system (Quantum Design).
%
\section
{
Results and discussion: Hexagonal DyMnO$_{3}$
}
\subsection
{
Crystal structure
}
%
As-grown crystals of both {\it h,o-}DyMnO$_{3}$ were black and
lustreless.
The XRD data for {\it h-}DyMnO$_{3}$ and results of the Rietveld
analysis are displayed in figure 1.
The crystal structure has been refined in a hexagonal space group
$P6_3cm$.
The refined lattice parameters are $a=$~6.189(1)~\AA~and
$c=$~11.461(4)~\AA.
Previous structural studies on {\it h-}$R$MnO$_{3}$ report the
structure in $P6_3cm$ space group with similar values for the
lattice parameters \cite{Zhouprb06}.
The indexed SAED pattern, figure 2, confirms the structure
refinement by the Rietveld method.
%
\subsection
{
{\it DC} Magnetisation
}
%
The field cooled (FC) as well as zero field cooled (ZFC)
magnetisation curves at a field of 10~Oe applied along the $c$
axis presented in figure 3(a) show bifurcation at about 3~K.
Similar data obtained for field  applied perpendicular to the $c$ axis
are shown in the inset of figure 3(a).
The temperature evolution of magnetisation is different from the properties
reported for {\it h-}HoMnO$_3$ or {\it h-}ScMnO$_3$ where such a
bifurcation at low temperature was not visible
\cite{Munozcm01,Munozprb00}.
It is interesting to note that the antiferromagnetic transition of
Mn at  $T^{\mathrm {Mn}}_N$~=~57~K, (determined from
specific heat measurements) is not discernible in the
magnetisation profile of {\it h-}DyMnO$_{3}$.
At temperatures above $T^{\mathrm {Mn}}_N$, the inverse
susceptibility 1/$\chi$ follows a Curie-Weiss behaviour, yielding
an effective magnetic moment value $\mu_{\mathrm
{obs}}=$~10.81~$\mu_B$(figure 4(b)).
This moment is close to the expected value
$\mu_{\mathrm {nom}}$ = $[\mu_{\mathrm {eff}}^2$(Mn)$ +
\mu_{\mathrm {eff}}^2$(Dy)$]^{1/2}=11.67~\mu_B$.
The Weiss temperature obtained from the fit is $\theta_W\approx
-23$~K.
The negative value of $\theta_W$ indicates the presence of
antiferromagnetic exchange interactions.
The values of $\mu_{\mathrm {obs}}$ and $\theta_W$ agree well with
those (10.7$\mu_B$ and $-$17~K) obtained for similar {\it
h-}HoMnO$_3$ crystals \cite{Munozcm01}.
However, as the magnetic response in the paramagnetic regime is
strongly influenced by excitations of the 4$f$-electrons of Dy
split by the crystalline electric field (CEF), the parameters from
the Curie-Weiss fit do not directly relate to the physics of the
coupled magnetic sublattices as in other hexagonal $R$MnO$_3$
systems \cite{Tomuta03}.

The isothermal magnetisation curves at 2, 10 and 60~K
measured along the $c$ axis and within the $ab$ plane are
presented in figure 4(a).
For $T \ge $~10~K the magnetisation and initial
susceptibility for fields applied  in the $ab$ plane
is larger than those measured along the $c$ axis.
This is consistent with (i) the easy-plane character of the
non-collinear magnetic order in the triangular Mn-sublattice, (ii)
the easy-axis anisotropy introduced by the Dy-ions along the
$c$-axis.
The data for 10~K and 60~K do not show any remanent magnetisation,
thus spontaneous or weak ferromagnetic moments are absent at these
temperatures.
The overall magnetisation increase for this temperature range is
probably dominated by a large linear contribution from the
paramagnetic Dy moments.
At 2~K, a ferromagnetic-like hysteretic behaviour is found in the
low-field part of the magnetisation curve (figure 4 (b)) and the
initial susceptibility is stronger along the $c$ axis than along
the $ab$ plane.
The magnetic behavior at this low temperature must be attributed
to the Dy-magnetic order and is strongly anisotropic.
The magnetisation curve along $c$ has the character of a
ferromagnetic easy-axis system with  technical saturation reached
at about 600~Oe.
In order to estimate the zero-field magnetic moment we
extrapolated the linear part of $M(H)$ above 600 Oe towards zero
field and obtained $M_S = $~7.5~emu/g~$ = $~0.36~$\mu_B/{\textrm{f.u.}}.$
Assuming that this spontaneous magnetisation is only due to a
ferromagnetic contribution of the Dy-moments on the 2(a) sites,
still the observed value for the ordered magnetic moment is very
small, only about 0.1 times the full polarization on the 2(a)-sites.
Hence, it is likely that the Dy-moments responsible for this
effect are only partially polarized by a coupling to the
Mn-sublattice and/or the 4(b)-sublattice, e.g. by dipolar
couplings.
Superimposed on this essentially ferromagnetic behaviour with a
small saturation magnetisation is a linear overall increase at low
fields 600~Oe~$< H <$~20~kOe and a tendency towards lower
high--field susceptibility, although no saturation is attained
even at 50~kOe.
Correspondingly, the magnetisation curve along the $ab$ plane
resembles a hard-axis magnetisation process with an essentially
linear increase of magnetisation.
However, the small hysteresis seen in the $ab$ magnetisation
curve cannot be explained by this process.
%
%
The existence of the different, yet coupled sublattices of Mn and
Dy, as mentioned above, underlies the high magnetic anisotropy and
the complex magnetisation behavior in this crystal.

A preliminary qualitative picture of the magnetic structure of
{\it h-}DyMnO$_{3}$ based on the magnetic measurements is possible
by a {\textquoteleft}three sub-lattice model{\textquoteright}
\cite{Tomuta03}.
This model assumes three different and largely decoupled magnetic
sublattices in {\it h-}$R$MnO$_3$: a lattice of Mn (at 6c) that
forms a 120$^{0}$ spin  arrangement in the $\it{ab}$ plane, and
two more lattices formed by the rare-earth ions $R1$ and $R2$
situated at 2(a) and 4(b) sites, respectively. A schematic of the
arrangement is shown in figure 5.

As the temperature is reduced, at around 57~K, the interlayer
coupling between the Mn ions in the $ab$ plane gives rise to a
non-collinear AF order, as found by specific heat data (see
below).
However, the exchange field originating from the Mn order below
$T^{\mathrm {Mn}}_N$ acts on the rare-earth lattice, and with
further reduction in temperature, the $R1$ and $R2$ moments
commence to order.
In the case of DyMnO$_{3}$, the Dy$^{3+}$ moments order below
$T^{\mathrm {Dy}}_N$~=~3~K.
At 2~K, a ferromagnetic feature is detected when measuring along
$c$ axis whereas along the $ab$ plane the ferromagnetic component
is weak.
In analogy to ErMnO$_3$ \cite{Tomuta03} this suggest that
the rare-earth ions at $R1$ could be paramagnetic down to low
temperatures before they order ferromagnetically.
On the other hand, the $R2$ ions may order antiferromagnetically
with their magnetic moments pointing along the hexagonal
$c$ axis similar to the magnetic order
observed in YbMnO$_3$ and TmMnO$_3$ \cite{Tomutajpcm01}.
With further increase in magnetic field, at about 30~kOe,
the bending in the $M(H)$ curves for both directions of
applied fields signals a gradual saturation at 2 and 10 K.
There is no clear explanation for this effect, however, it should
be noted that a contribution of the Mn-sublattice is likely, as a
similar bending of $M(H)$ curves has been observed for {\it
h-}YMnO$_3$ with a non-magnetic $R$-site at low temperatures
\cite{Munozprb00}.
%

%
\subsection
{
{\it AC}-Susceptibility
}
%
Recently, observations on thin films of {\it h-}DyMnO$_3$
indicated that the competition and frustration inherent in the
magnetic multi-sublattice of $R$MnO$_3$ may lead to a magnetic
spin-glass state \cite{Leeapl07}.
However, from the {\it ac}-susceptibility of {\it h-}DyMnO$_{3}$
along $c$ and $ab$ the transition of the Mn lattice is not obvious
and possibly being masked by the frustration in the Mn lattice or
by the strong Dy paramagnetic susceptibility.
It is clear that the competing inter- and intra-sublattice
couplings are present in the hexagonal structure of $R$MnO$_3$
which can have pronounced effects at low temperatures.
As expected, the real part of the susceptibility, $\chi'$, exhibits
a peak close to 3~K which is evident in the $\chi'(T)$ curves of
{\it h-}DyMnO$_3$ for the range of frequencies 133~Hz -- 10~kHz
measured along the $c$ axis as well as the $ab$ plane, figure 6(a)
and (b), respectively.
This peak originates from the ordering of Dy moments at low
temperature.
The peak temperature displays no significant dependence on
frequency. This rules out the presence of slow dynamics as a
characteristics of glassy magnetism.

Figure 7(a) demonstrates the effect of magnetic field on $\chi'$
for fields up to 70~kOe. With the application of a magnetic field
along $c$-axis, the antiferromagnetic signal is suppressed and the
peak at 3~K diminishes in magnitude.
Above 30~kOe, the peak vanishes corroborating a gradual
field-driven transformation of the magnetic structure in this
field range.
When a magnetic field is applied along the $ab$ plane (figure
7(b)), one observes a shift in the peak to higher temperatures,
in addition to a decrease in the magnitude of $\chi'$.\\
%
\subsection
{
Specific heat
}
%
The specific heat $C_p$ of {\it h-}DyMnO$_{3}$ measured at zero
applied field is presented in figure 8 (inset magnifies the two
apparent transitions).
The antiferromagnetic transition of the Mn sublattice is evident
as a sharp peak at about 57~K.
Thus, in {\it h-}DyMnO$_3$ the magnetic ordering transition in the
Mn-sublattice appears only in specific heat, but is not detectable
in the magnetic susceptibility data.
This is in contrast to the properties of hexagonal HoMnO$_3$,
where $T_{N}^{\mathrm {Mn}}$ is detected by peaks in $C_p$  and in
$\chi ''$ as well \cite{Munozcm01}.
In contrast to the specific heat of {\it h-}HoMnO$_3$ studied by
Mu\~{n}oz $et~al.$ \cite{Munozic01}, no sign of a Mn reorientation
in the basal plane at intermediate temperatures is observed in
{\it h-}DyMnO$_3$.
At 3~K, a second peak is observed in the specific heat
corresponding to the ordering of Dy-moments.
%
\section
{
Results and discussion: Orthorhombic DyMnO$_{3}$
}
\subsection
{
Crystal structure
}
%
Generally, the {\it o-}$R$MnO$_3$ derives from the perovskite
structure \cite{Kimuraprb05}.
The powder XRD data for {\it o-}DyMnO$_{3}$ used for the
structural analysis are displayed in figure 9 along with the
Rietveld refinement in the $Pnma$ space group.
The lattice parameters obtained are a = 5.832(1)~\AA, b =
7.381(2)~\AA~ and c = 5.280(1)~\AA, respectively.
Earlier structural studies \cite{Moriml00} on single crystalline
DyMnO$_3$ reported similar values.
The structure solution obtained for {\it o-}DyMnO$_3$ is also
supported by the SAED pattern, figure 10, which could be indexed
in the orthorhombic structure.
The perovskite DyMnO$_3$ presents a highly distorted structure
owing to the small value of the ionic radius $r_R$ of Dy at the
$R$ site, the tolerance factor of DyMnO$_3$ being about 0.85.
The lattice distortions -- Jahn Teller distortion and octahedral
rotation -- observed in perovskite $R$MnO$_3$ evolve continuously
with decreasing $r_R$ \cite{Tachibanaprb07}.
Nevertheless, such a smooth evolution of distortions with $r_R$
does not imply a similarly continuous evolution of the magnetic
structure \cite{Zhouprl06}.
Normally, a reduction in $r_R$ is followed by a deviation of the
Mn--O--Mn bond angle from 180$^{\circ}$, thereby distorting the
major magnetic exchange path. In turn, the magnetic structure of
$R$MnO$_3$ changes from $A$ type to $E$ type as a function of
$r_R$ \cite{Tachibanaprb07,Zhouprl06}.
However, smaller rare-earths like Dy, Tb and Gd fall in the
intermediate region of bond angle values that stabilize
incommensurate magnetic structures \cite{Kimuraprb03}.
%
\subsection
{
{\it DC} Magnetisation
}
%
The Mn magnetic moments in the perovskite DyMnO$_{3}$ are directed
along the orthorhombic $b$-axis \cite{Prokhnenkoprl07}.
Below $T_N$ $\approx$ 40~K the Mn spins enter the incommensurate
state where they order sinusoidally with modulation vector along
(0, q$_{\mathrm {Mn}}$, 0) \cite{Yeprb07,Kajimotoprb04}.
Based on neutron scattering measurements \cite{Kimuranature03} it
is further reported that the sinusoidal order transforms into an
incommensurate spiral order through a {\textit{lock--in}}
transition at $T_{\mathrm{lock-in}}$ = 18~K.
At even lower temperature the rare-earth spins also order,
$T^{\mathrm {Dy}}_N\,\approx 6$~K.

The magnetisation curves obtained in FC and ZFC cycles
with an applied field of 20~Oe parallel to $b$ axis
are presented in figure 11(a).
Three features can be recognized from these curves; a bifurcation
in FC/ZFC cycles at $T_{\mathrm{split}} \approx$ 45~K (magnified in 
the second inset for clarity), a peak in
the ZFC curve at 16 K, and an inflection in the FC magnetisation
curve at 5~K.
These features can be attributed to the incommensurate Mn-spin
order, to the {\it lock--in} transition, and to the Dy-spin order,
respectively.
Values of $T^{\mathrm {Dy}}_N$ at 9 and 6~K have been reported for
polycrystalline \cite{Kimuraprb03} and single crystal samples of
DyMnO$_3$, respectively \cite{Prokhnenkoprl07}.
At higher fields, the ZFC magnetisation curve exhibits a peak at
lower temperature around 7~K, figure 11(b).
However, no signature of the ordering transition into the
incommensurate structure at 39~K is visible.
Interestingly, the low temperature rise in ZFC magnetisation
persists even in an applied field of 1000~Oe.
The 1/$\chi$ data (not shown) above $T_N$ follow a Curie-Weiss
behaviour yielding a value of 13.67~$\mu_B$ for $\mu_{obs}$. This
is higher than the expected value $\mu_{nom}$ = 11.67~$\mu_B$
calculated from $\mu_{\mathrm {nom}}$ = $[\mu_{\mathrm
{eff}}^2$(Mn)$\, + \,\mu_{\mathrm {eff}}^2$(Dy)$]^{1/2}$.
From the fit, $\theta_W$~$\approx - 20$~K which again indicates the
presence of antiferromagnetic interactions in the system.
However, as in the case of {\it h}-DyMnO$_3$, the parameters of
the Curie-Weiss law do not reflect the properties of a simple
paramagnetic and coupled system since they are affected by the
transitions between the CEF-levels of Dy.

The low temperature behaviour of the magnetisation measured
perpendicular to the $b$ axis is substantially different from that
parallel to $b$, as can be inferred from figure 12(a) and (b).
Here, the FC/ZFC curves at 10 and 1000~Oe, respectively, are
presented.
Even though the bifurcation signalling the first transition into
the incommensurate structure and the {\it lock--in} transition are
evident (see also the insets of figure 12 (a)), both the FC and ZFC curves show a decrease of the
magnetisation below the peak at $\approx$ 7~K.
This is different from the low temperature behaviour of the
magnetisation parallel to $b$ and indicates that the magnetic easy
axis of the orthorhombic system is along the $b$ axis.
Although the absolute magnitude of magnetisation is higher in this
case, the difference between FC and ZFC is less pronounced.

Isothermal magnetisation curves at 2, 10 and 60~K with applied
field parallel to the $b$ axis are presented in figure 13.
No hysteresis is discernible and the magnetisation varies linearly
with applied field.
In the related compound HoMnO$_3$, Mu\~{n}oz $et~al.$ found a
metamagnetic transition at 2~K for fields higher than 50~kOe
\cite{Munozic01}.
In {\it o-}DyMnO$_3$, no such feature is observed at 2~K.
It is interesting to note that the ordering temperature of the
same rare-earth ion in {\it h-}DyMnO$_3$ (3 K) and {\it
o-}DyMnO$_3$ (7 K) are separated by about 4 K.
%
\subsection
{
{\it AC}-Susceptibility
}
%
The dependence of $\chi'$ on temperature at different probing
frequencies for {\it o-}DyMnO$_3$ is presented in figure 14(a).
The measurement was performed with applied field parallel to the
$b$ axis.
Although a weak dependence of $\chi'$ on frequency
is observed, the shift in the peak temperature $T_{peak}$
is very small ($\sim$ 1 K).
This peak corresponds to the $T_{\mathrm{lock-in}}$.
Both the incommensurate transition of the Mn-spins and the
antiferromagnetic transition of Dy spins are not visible in the
{\it ac}-susceptibility.
Application of a magnetic field suppresses this magnetic peak and,
at the same time, shifts $T_{peak}$ to lower temperature, 
figure 14(b).
The change of the low-temperature peak in $\chi'$ possibly
indicates a metamagnetic transition in the Dy spins.
A non-collinear arrangement can give rise to additional
ferroelectric displacements through magnetoelastic couplings
\cite{Mostovoyprl06}.

\subsection{Specific heat}
The heat capacity $C_p$ of {\it o-}DyMnO$_3$ measured
in zero field is shown in figure 15.
A sharp peak at about 39~K marks the antiferromagnetic
transition into the sinusoidal incommensurate phase.
Additional anomalies in $C_p$ are observed at
16~K and 5~K which, again, indicate the {\it lock--in} transition
and $T^{\mathrm {Dy}}_N$.
Therefore, the specific heat data corroborate
the main features of the phase diagram and
magnetic ordering in {\it o-}DyMnO$_3$.
%
\section
{
Conclusions
}

%
We have successfully synthesized the multiferroic crystal
DyMnO$_3$ in the much studied orthorhombic symmetry as well as in
the less studied hexagonal symmetry by employing different growth
ambiences.
The basic magnetic properties were investigated by means of
{\it dc} magnetisation, {\it ac}-susceptibility, and specific heat.\\
\textit{$h-$DyMnO$_{3}$:} \\
Hexagonal DyMnO$_3$ shows an antiferromagnetic transition at 57~K
clearly discernible in the specific heat measurement.
The huge paramagnetic susceptibility stemming from Dy and/or the
frustrated Mn lattice could be the reason for this transition not
being visible in the magnetization and susceptibility curves.
The rare-earth moment manifests itself through a partial
antiferromagnetic order in the FC/ZFC curves at 3~K.
The mixed interactions and frustration in the magnetic lattice do
not lead to a magnetic glassy state, as is inferred from the
absence of any frequency dependence in $\chi'$.
On application of a magnetic field the antiferromagnetic signal in $\chi'$ at
low temperature is suppressed.
In addition to the transition in the Mn sublattice the specific
heat data exhibit a peak at even lower temperature resulting from the
rare-earth magnetic ordering.
No sign of a Mn-reorientation transition is detected in the
specific heat measurement.\\
\textit{$o-$DyMnO$_{3}$:}\\
{\it DC} magnetisation measurements parallel and perpendicular to the
$b$ axis markedly differ in the case of {\it o-}DyMnO$_3$ from 
its hexagonal counterpart.
Parallel to $b$, the transition into the incommensurate phase
is observed in the FC and ZFC curves. These curves
split at 39~K, exhibit a peak in the ZFC magnetisation
curve at 16~K, and a weak
feature at 5~K the latter signalling the Dy order.
Perpendicular to $b$, peaks are observed both in the FC/ZFC
magnetisation curves at 7~K.
{\it AC} susceptibility measurements do not show
any signatures of the incommensurate transition
or any frequency dependence.
However, they clearly reflect the {\textit{lock--in}} transition.
The specific heat curves corroborate these conclusions displaying
peaks at temperatures corresponding to the sinusoidal ordering of Mn moments
into an incommensurate phase, the {\it lock--in} transition,  and
the  ordering of the rare-earth moments. \\
Our detailed magnetisation measurements
highlight the complex interplay of
the 3$d$ and rare-earth magnetism in $h$- and $o$-DyMnO$_3$.
In turn, investigating the effect of rare-earth magnetism of Dy on
the magnetic ordering and dielectric property of {\it o-}DyMnO$_3$
at low temperature is called for.

\section*{Acknowledgment}
We thank G. Behr (IFW Dresden) for help with single-crystal
orientation. The authors acknowledge the Department of Science and Technology
(DST), India for OFZ crystal growth facility set up through FIST programme.
SE and HLB thank DST for financial support through a project grant.
%


\clearpage
\begin{figure}
\begin{center}
\includegraphics[scale=1.5]{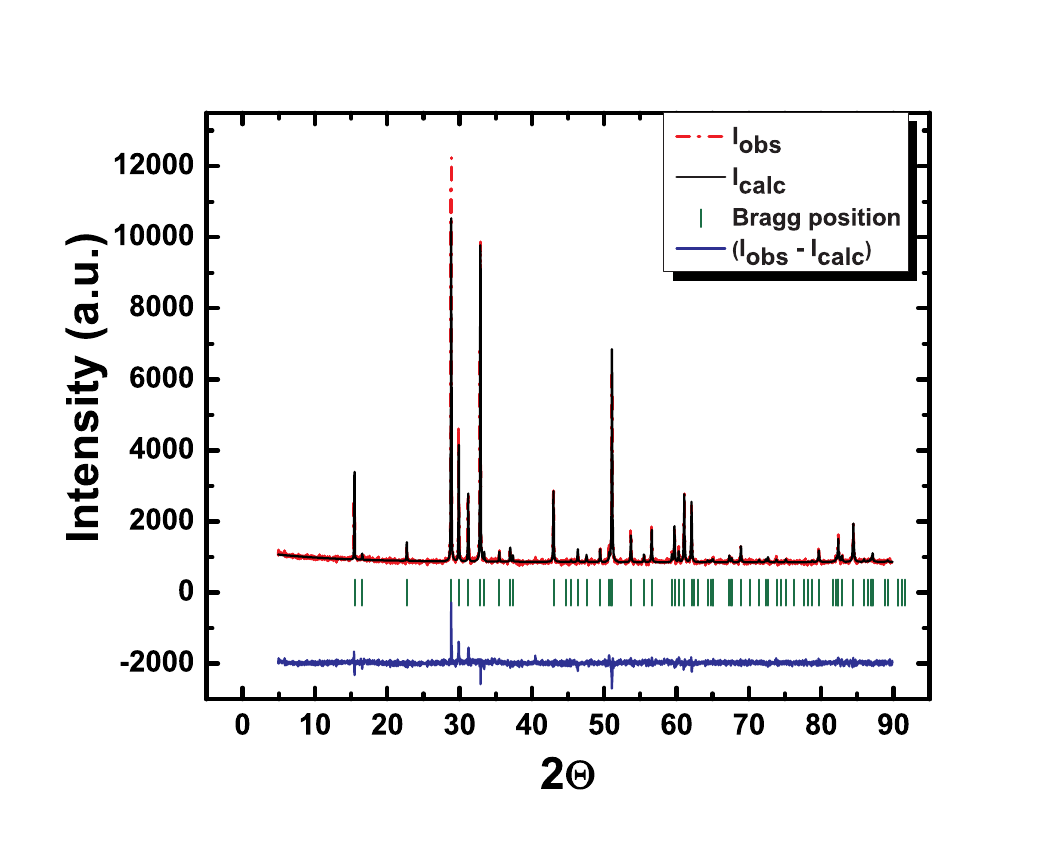}
\caption{Powder X-ray diffraction pattern ($I_{obs}$) and Rietveld
refinement ($I_{calc}$) of $\textit{h-}$DyMnO$_{3}$.}
\end{center}
\end{figure}

\clearpage
\begin{figure}
\begin{center}
\includegraphics[scale=1.5]{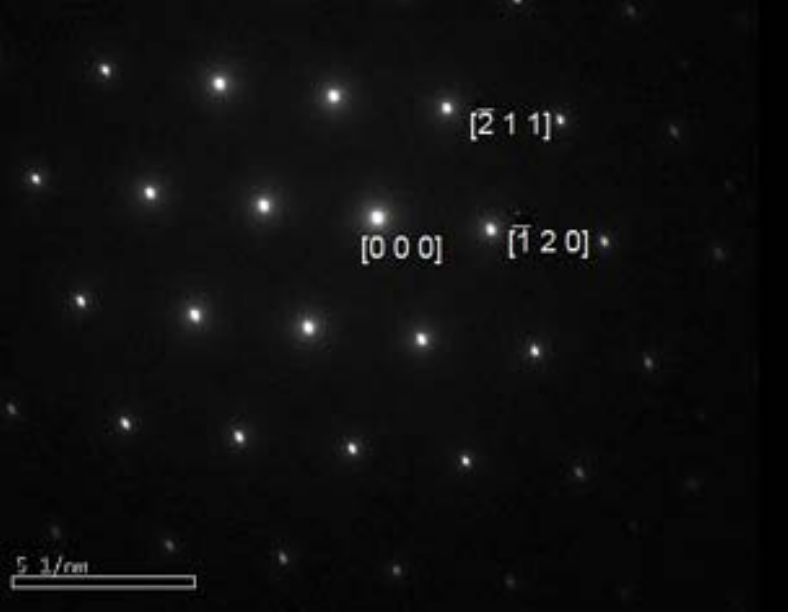}
\caption{SAED pattern of $\textit{h-}$DyMnO$_{3}$. The scale bar
shows 5 nm$^{-1}$.}
\end{center}
\end{figure}

\clearpage
\begin{figure}
\begin{center}
\includegraphics[scale=1.5]{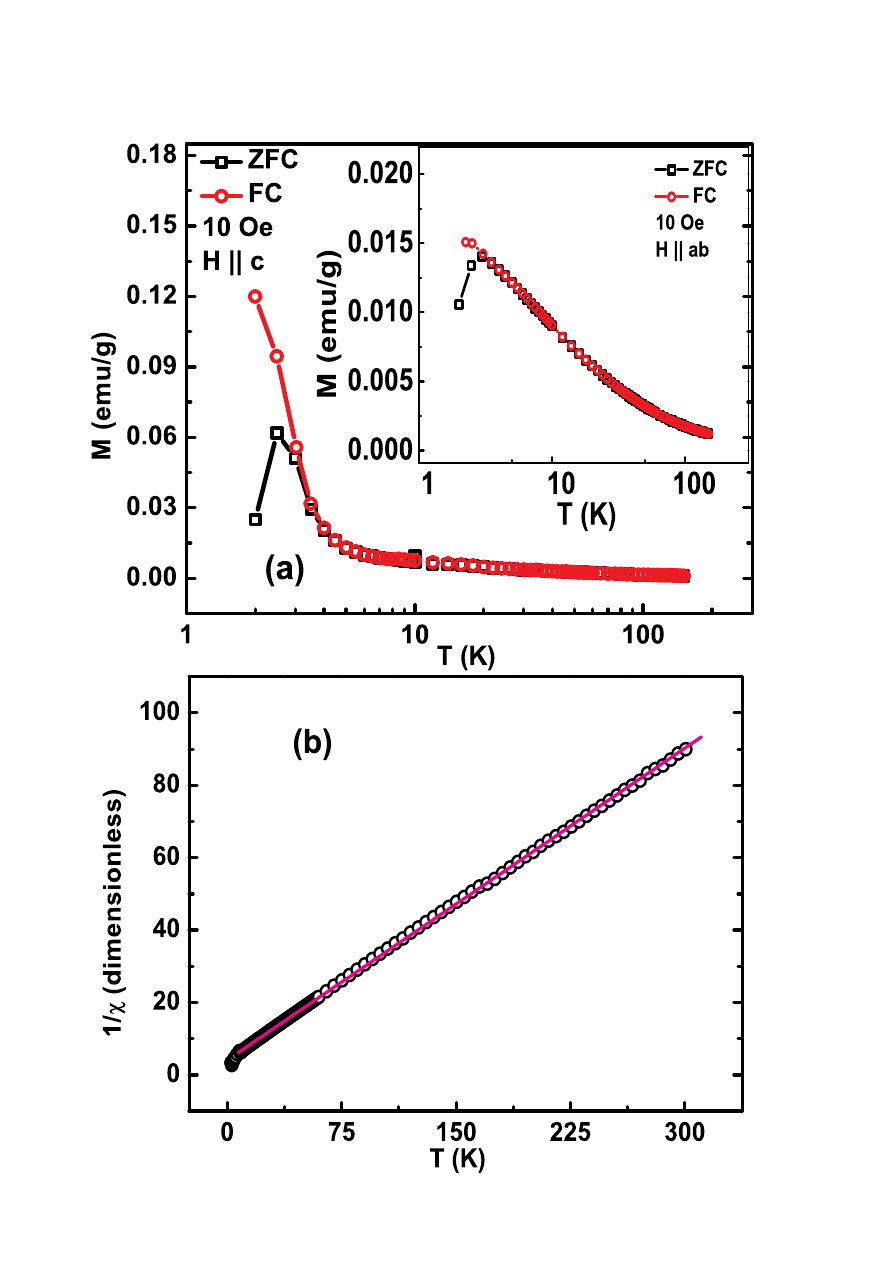}
\caption{(a) Magnetisation of $\textit{h-}$DyMnO$_{3}$
along the $\it{c}$ axis and the $\it{ab}$ plane (inset). The value
of magnetisation is an order of magnitude higher along the c-axis.
(b) A Curie-Weiss law fits the inverse susceptibility data
with a negative Weiss temperature and deviations at lowest
temperatures.}
\end{center}
\end{figure}

\clearpage
\begin{figure}
\begin{center}
\includegraphics[scale=1.5]{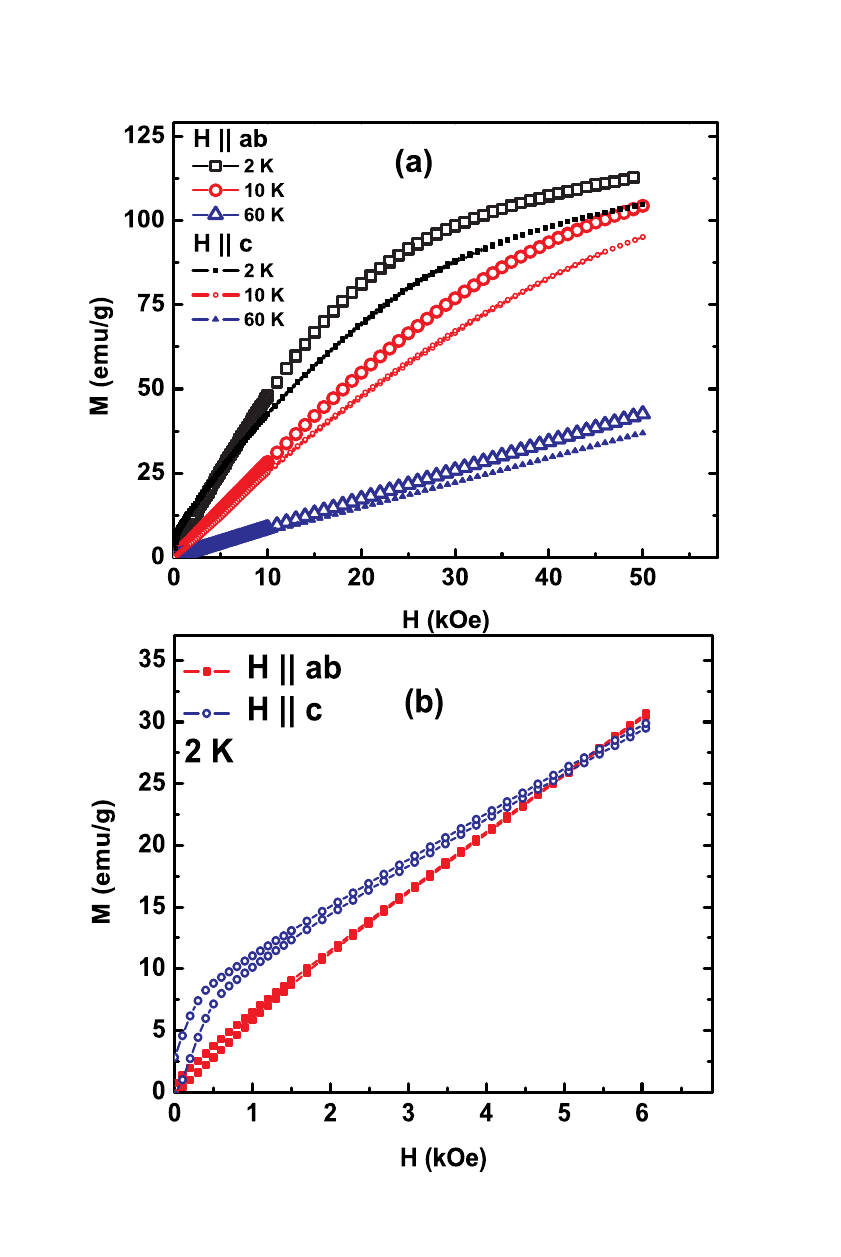}
\caption{(a) Magnetisation isotherms of $\textit{h-}$DyMnO$_{3}$
at 2, 10 and 60~K measured along $\it{c}$ and $\it{ab}$
directions. (b) Low-field magnetisation along $\it{c}$ showing a
weak spontaneous magnetic moment and hysteretic behaviour.}
\end{center}
\end{figure}

\clearpage
\begin{figure}
\begin{center}
(a)\includegraphics[scale=0.35]{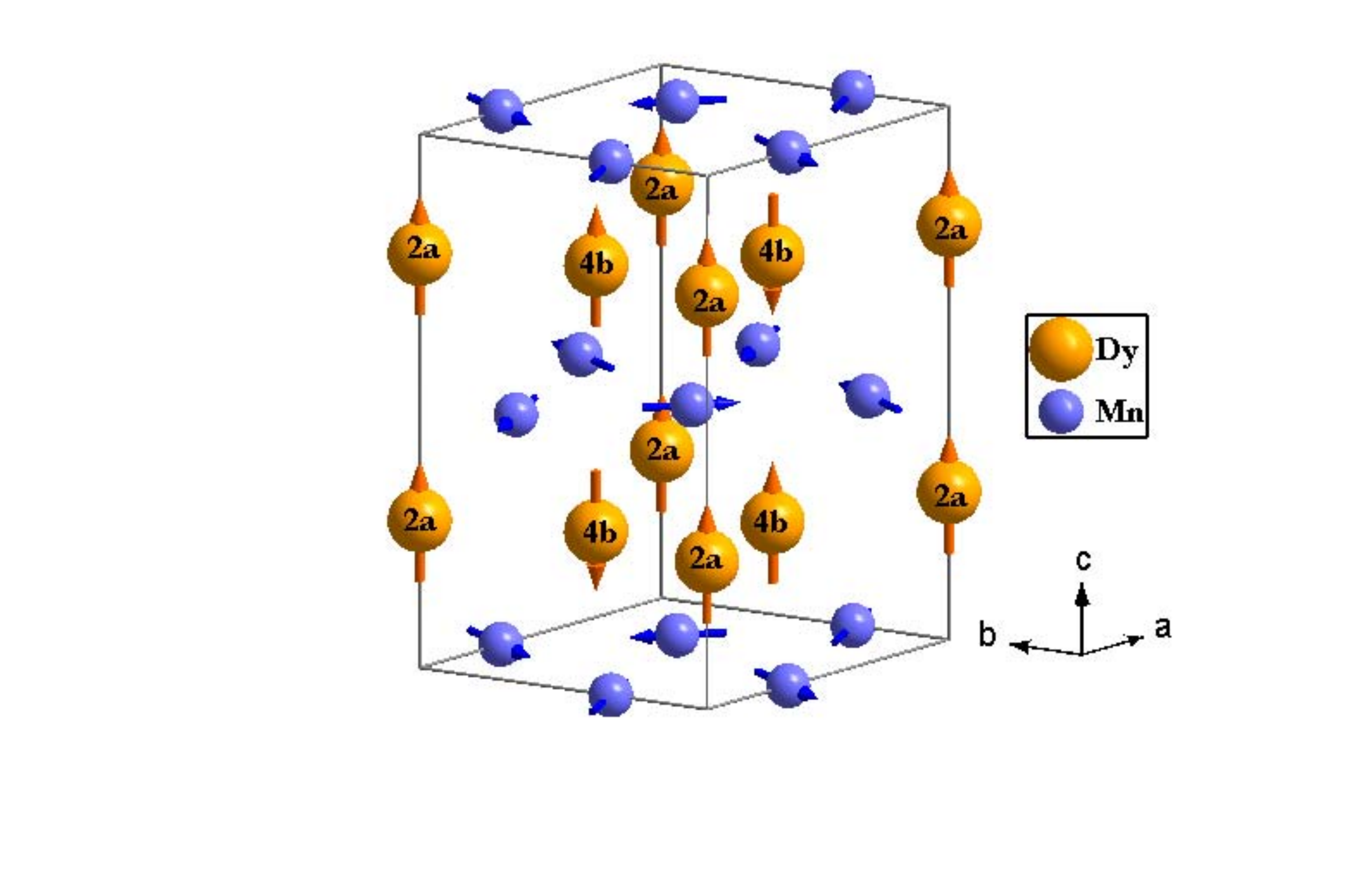}
(b)\includegraphics[scale=0.35]{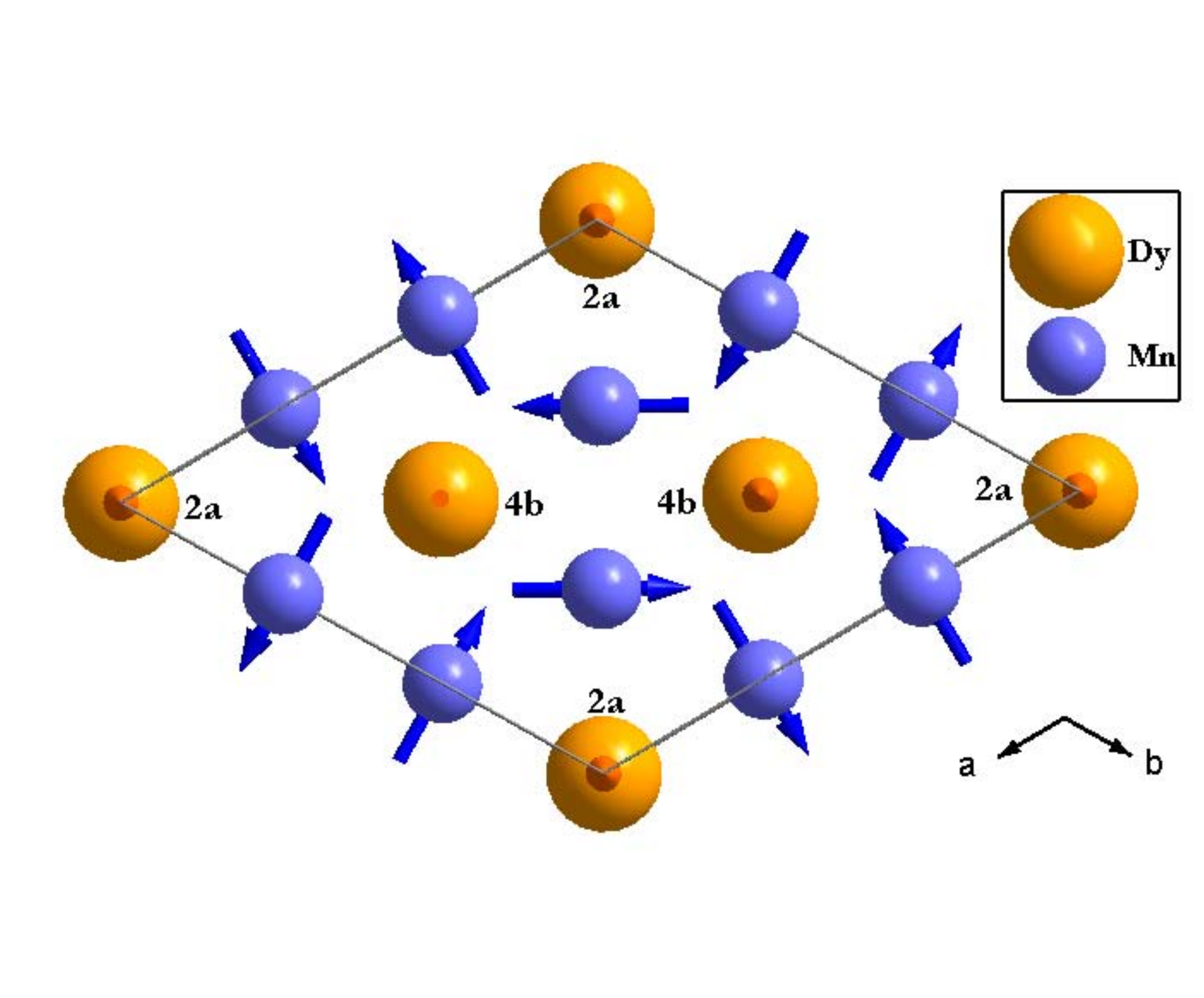}
(c)\includegraphics[scale=0.35]{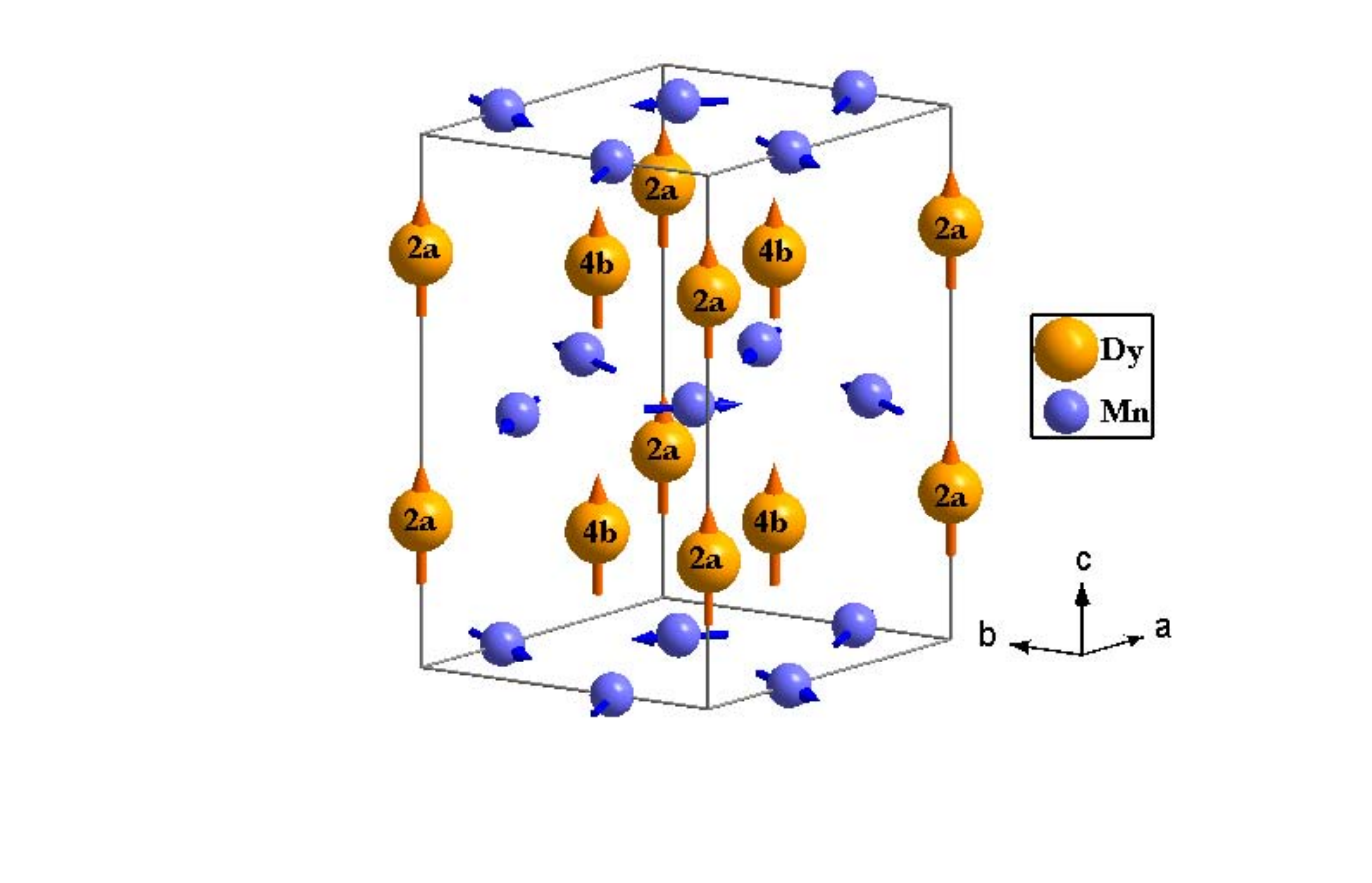} \caption{Magnetic
structure of the hexagonal DyMnO$_3$ crystal. The Dy$^{3+}$ ions
at the two inequivalent positions along with the Mn ions are shown.
Note that, although the moments at 2(a) sites are depicted as
ferromagnetically aligned, the spins at this site are only
partially polarized. (a) and (b) display the zero-field case while
(c) presents the magnetic lattice above 30~kOe.}
\end{center}
\end{figure}

\clearpage
\begin{figure}
\begin{center}
\includegraphics[scale=1.5]{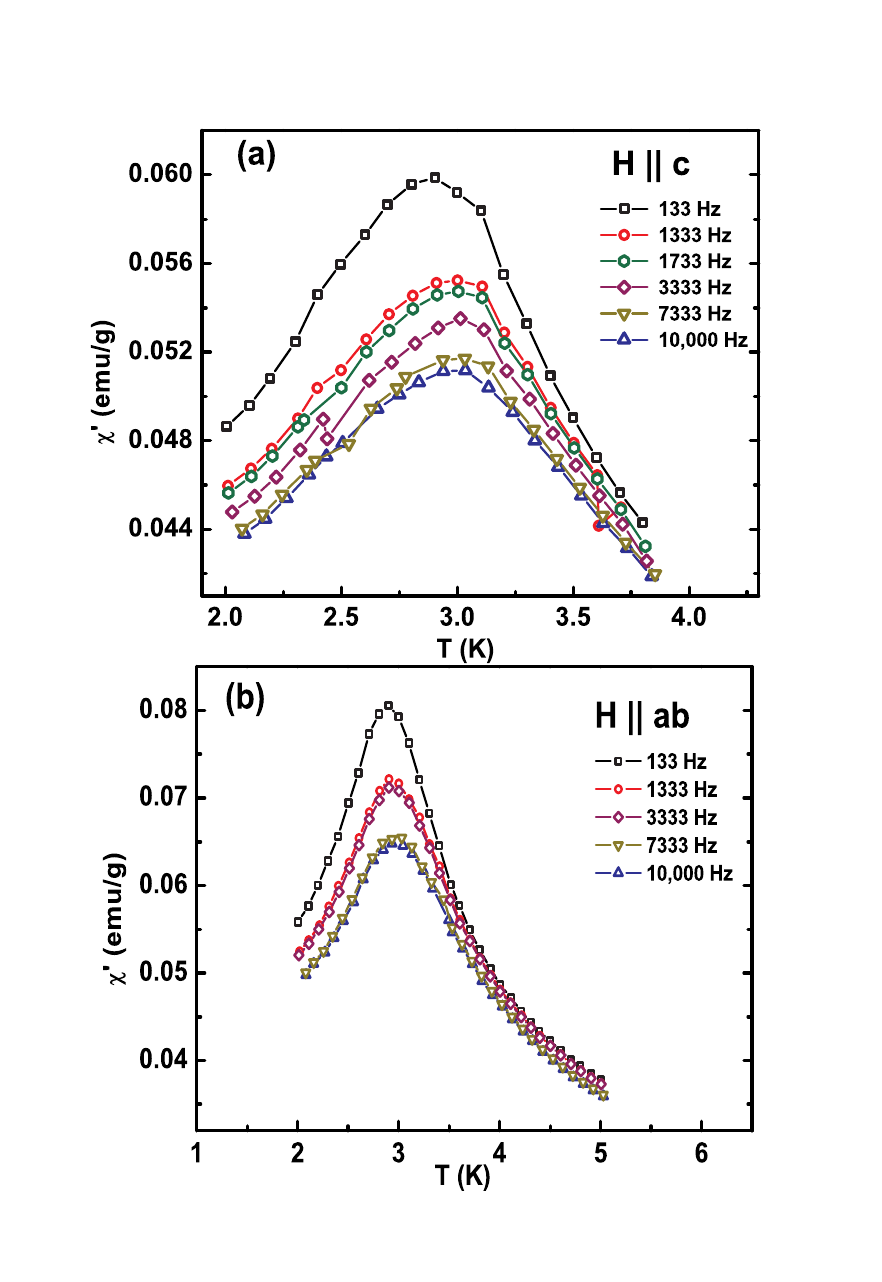}
\caption{{\it AC} susceptibility curves for
$\textit{h-}$DyMnO$_{3}$ with magnetic field $H$ along (a) the
$\it{c}$ axis and (b) the $\it{ab}$ plane at different
frequencies. The amplitude of the ac field was 10~Oe. 
There is only a very feeble dependence of the peak
temperature on frequency.}
\end{center}
\end{figure}

\clearpage
\begin{figure}
\begin{center}
\includegraphics[scale=1.5]{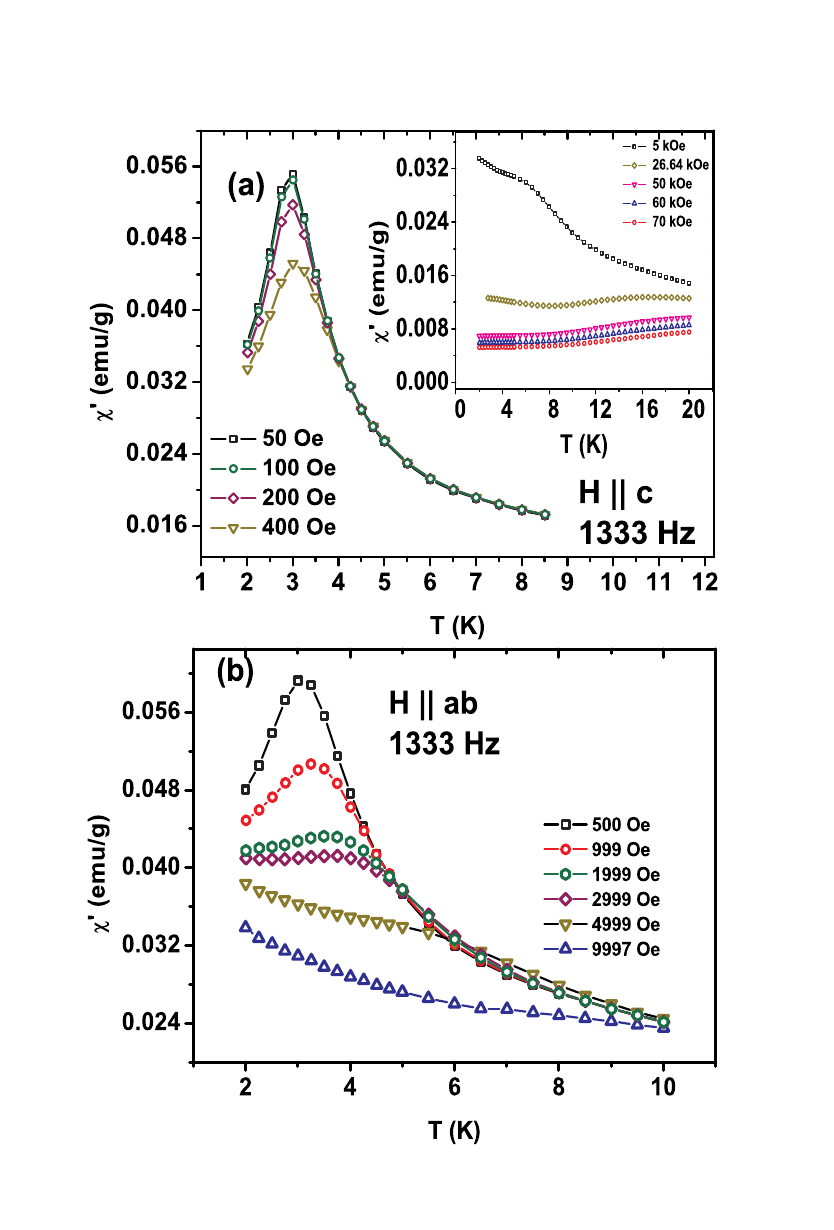}
\caption{Effect of an imposed magnetic field $H$ on the susceptibility of
$\textit{h-}$DyMnO$_{3}$ with $H$ along (a) the
$\it{c}$ axis and (b) the  $\it{ab}$ plane. 
The amplitude of the ac field was 10~Oe.}
\end{center}
\end{figure}

\clearpage
\begin{figure}
\begin{center}
\includegraphics[scale=1.5]{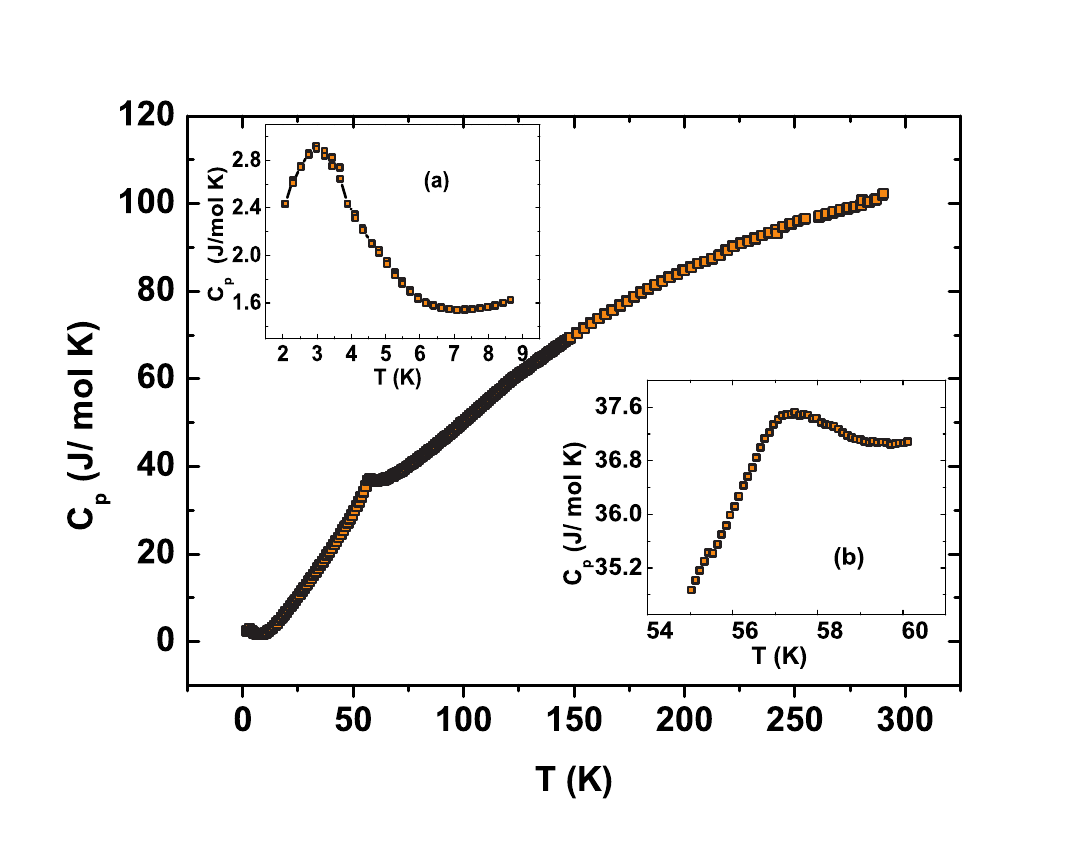}
\caption{Specific heat at zero applied field for
$\textit{h-}$DyMnO$_{3}$. The insets magnify (a) the low
temperature peak and (b) the peak at the Mn antiferromagnetic
transition.}
\end{center}
\end{figure}

\clearpage
\begin{figure}
\begin{center}
\includegraphics[scale=1.5]{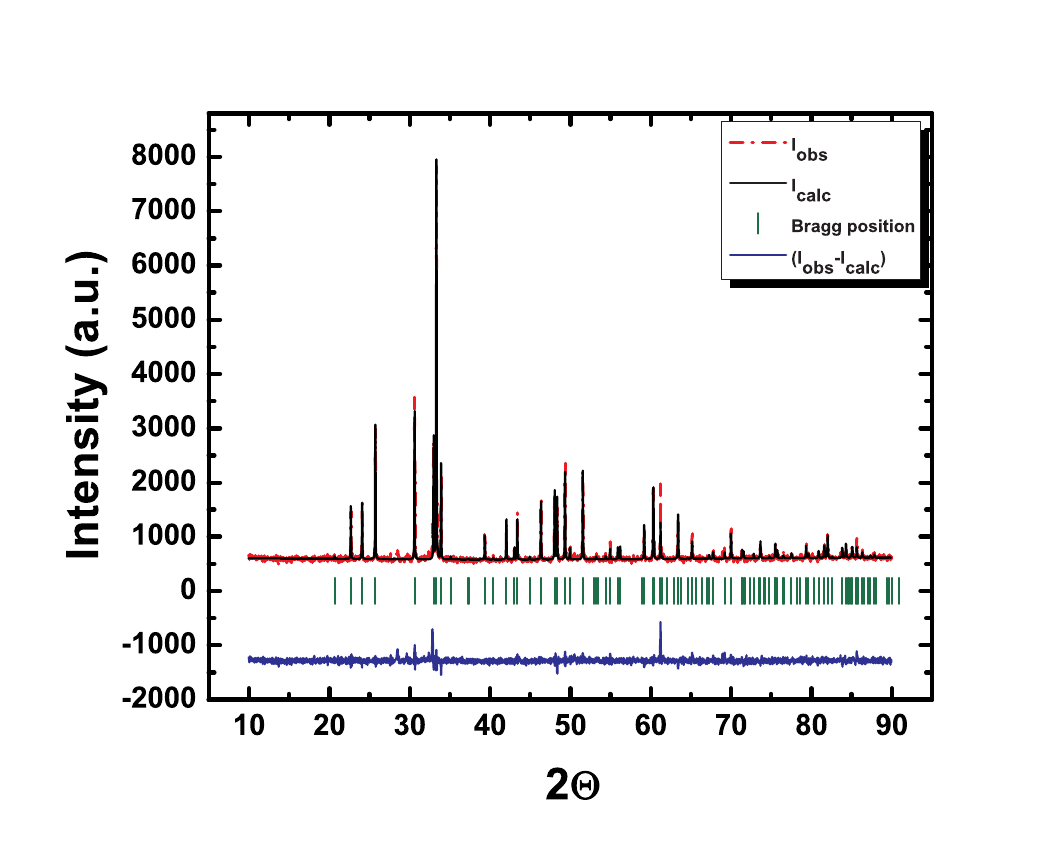}
\caption{The powder X-ray diffraction pattern
($I_{obs}$) and the Rietveld refinement ($I_{calc}$) of
$\textit{o-}$DyMnO$_{3}$.}
\end{center}
\end{figure}

\clearpage
\begin{figure}
\begin{center}
\includegraphics[scale=0.8]{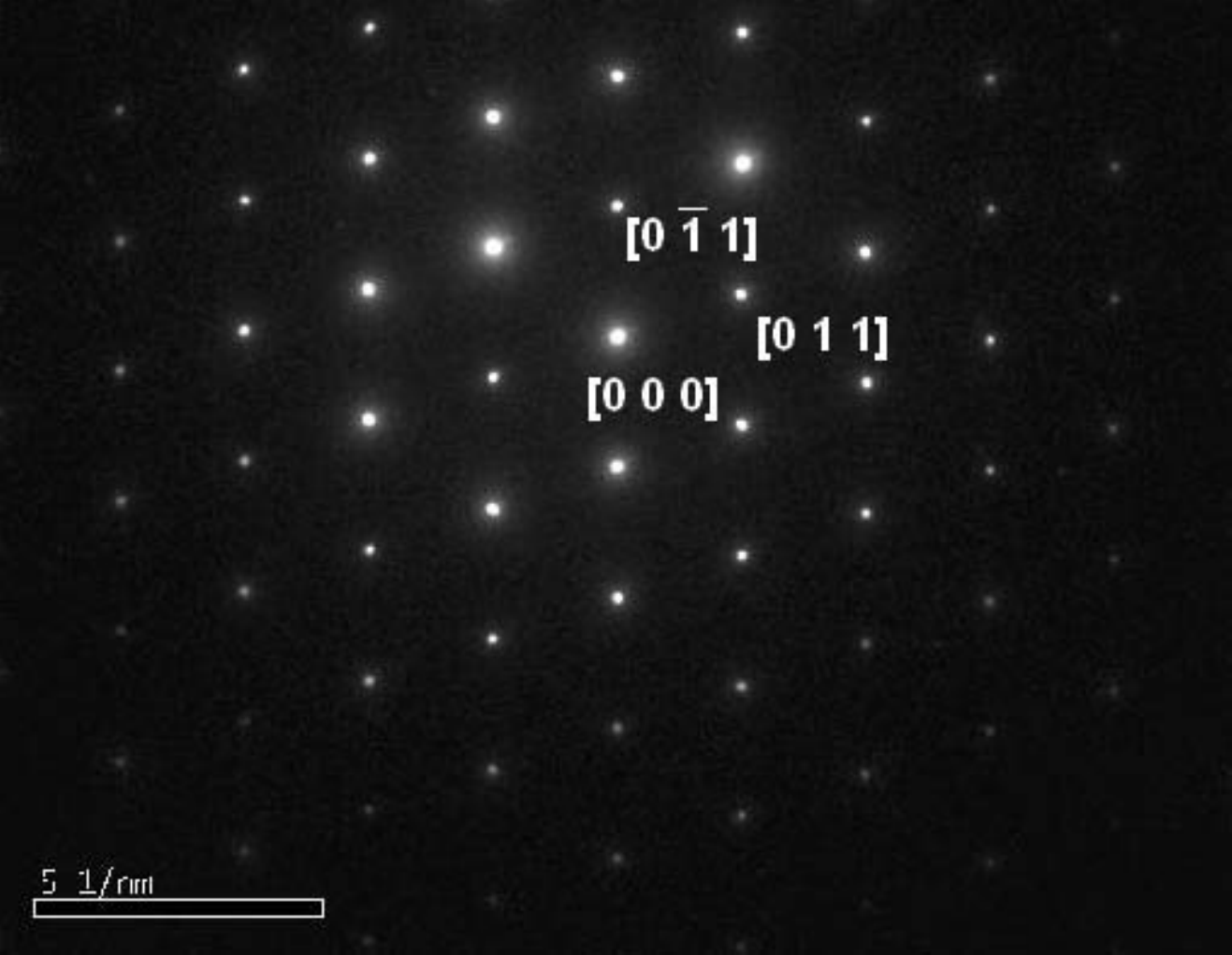}
\caption{SAED pattern of $\textit{o-}$DyMnO$_{3}$. The scale bar
shows 5 nm$^{-1}$.}
\end{center}
\end{figure}

\clearpage
\begin{figure}
\begin{center}
\includegraphics[scale=1.5]{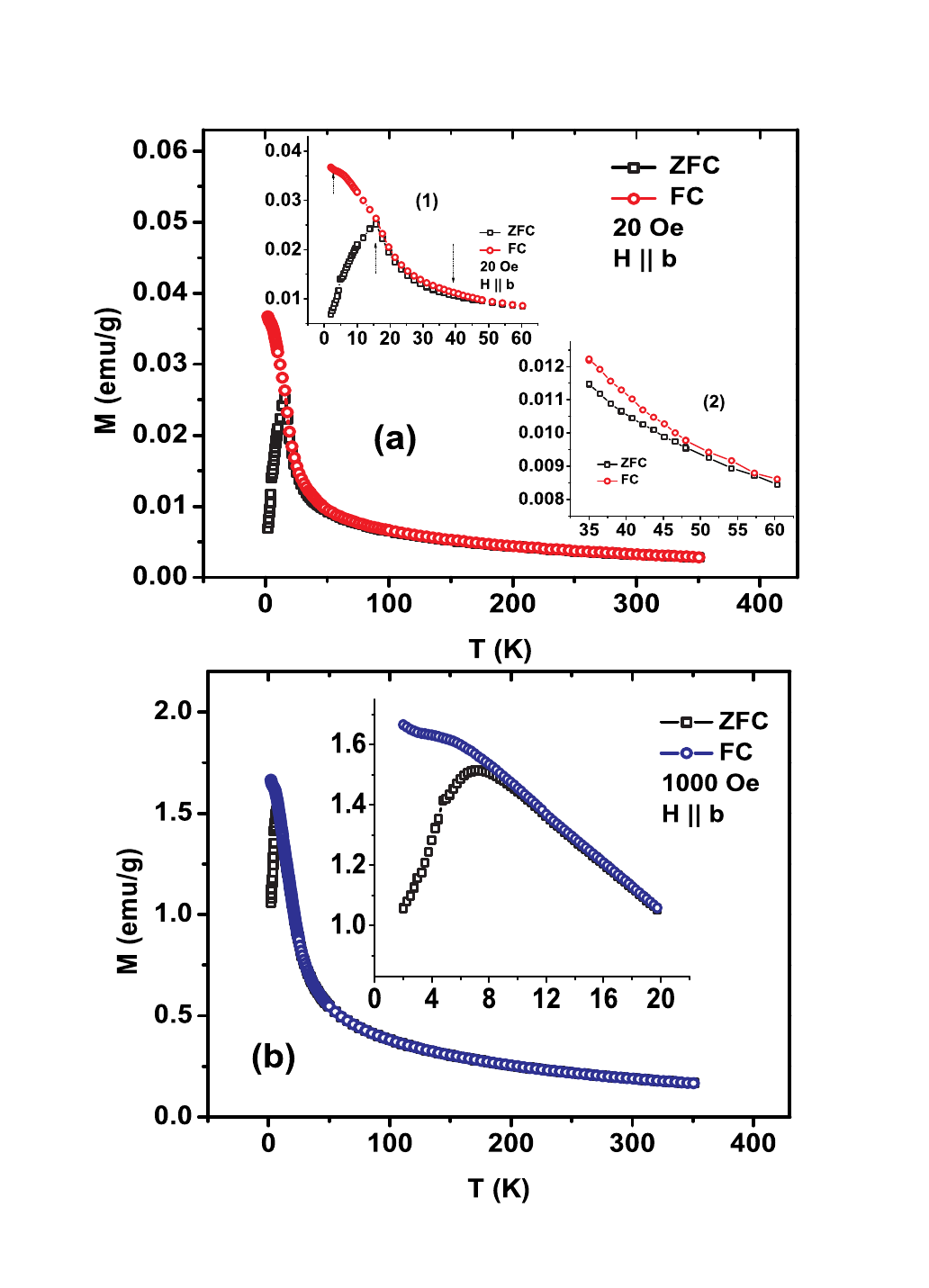}
\caption{Magnetisation in dependence on temperature of
$\textit{o-}$DyMnO$_{3}$ $parallel$ to $\it{b}$~axis with (a) $H$
= 20~Oe and (b) $H$ = 1000~Oe. The three transitions described in the
text are indicated by arrows in the inset 1 of (a). Inset 2 of (a) magnifies the ZFC-FC bifurcation around 45 K.
The peak in the ZFC magnetization shifted to lower temperatures in higher applied field can be clearly seen in the inset of (b).}
\end{center}
\end{figure}

\clearpage
\begin{figure}
\begin{center}
\includegraphics[scale=1.5]{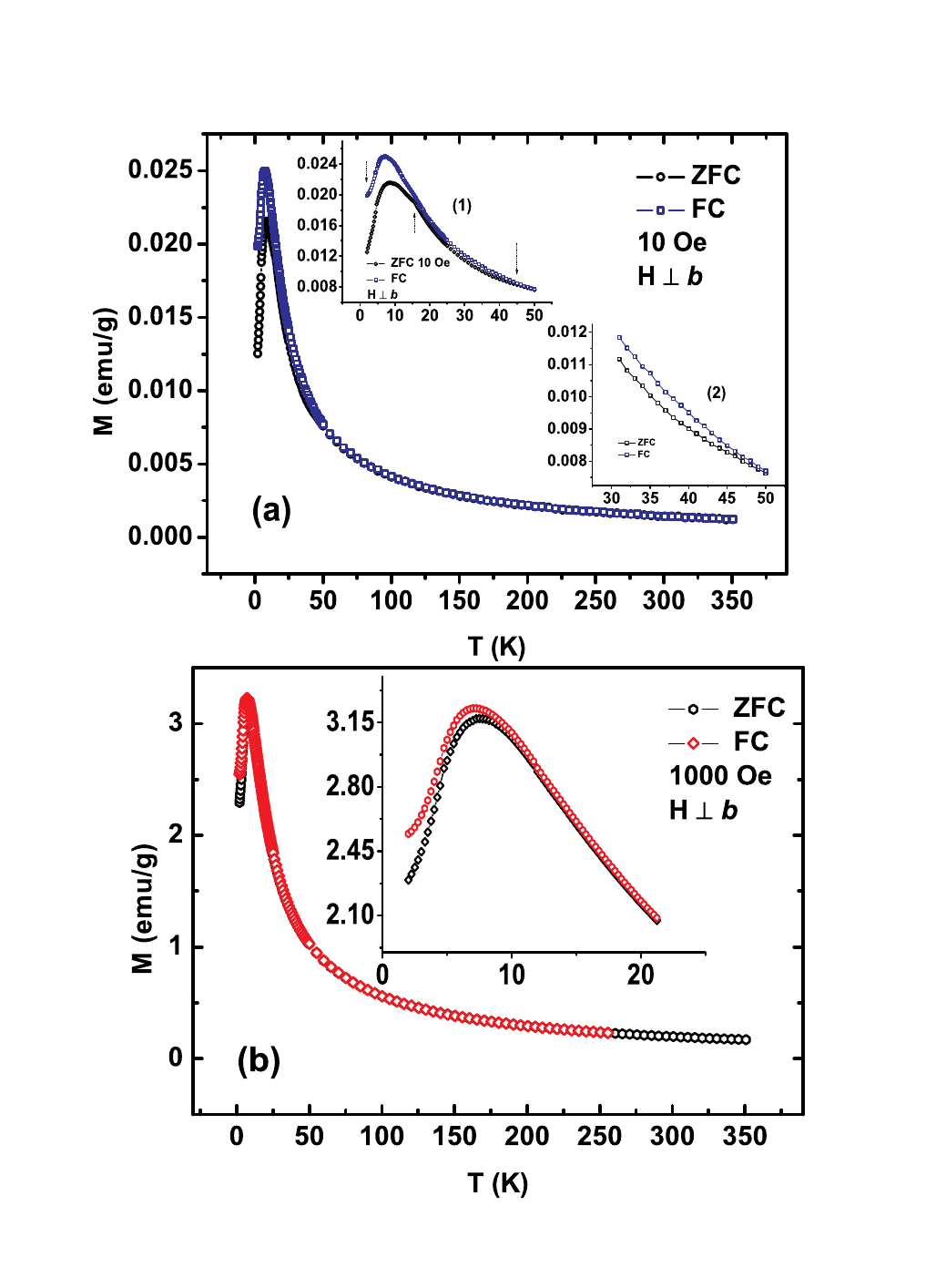}
\caption{Magnetisation of $\textit{o-}$DyMnO$_{3}$ $perpendicular$
to $\it{b}$~axis with (a) $H$ = 10~Oe and (b) $H$ = 1000~Oe. The
three transitions described in the text are indicated by arrows in
the inset 1 of (a). Inset 2 of (a) magnifies the ZFC-FC bifurcation around 45 K. Inset of (b) magnifies the magnetization with $H$ = 1000~Oe at temperatures below 20 K.}
\end{center}
\end{figure}

\clearpage
\begin{figure}
\begin{center}
\includegraphics[scale=1.5]{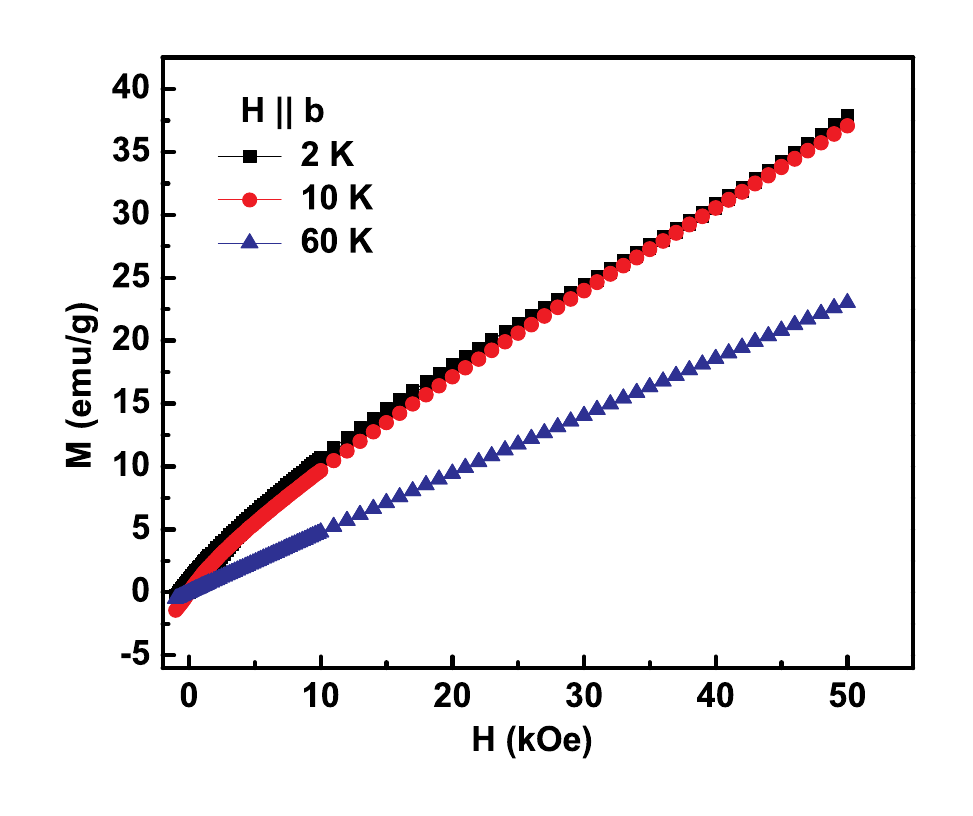}
\caption{Magnetisation of $\textit{o-}$DyMnO$_{3}$ at
2, 10 and 60~K with magnetic field parallel to $b$-axis}
\end{center}
\end{figure}

\clearpage
\begin{figure}
\begin{center}
\includegraphics[scale=1.5]{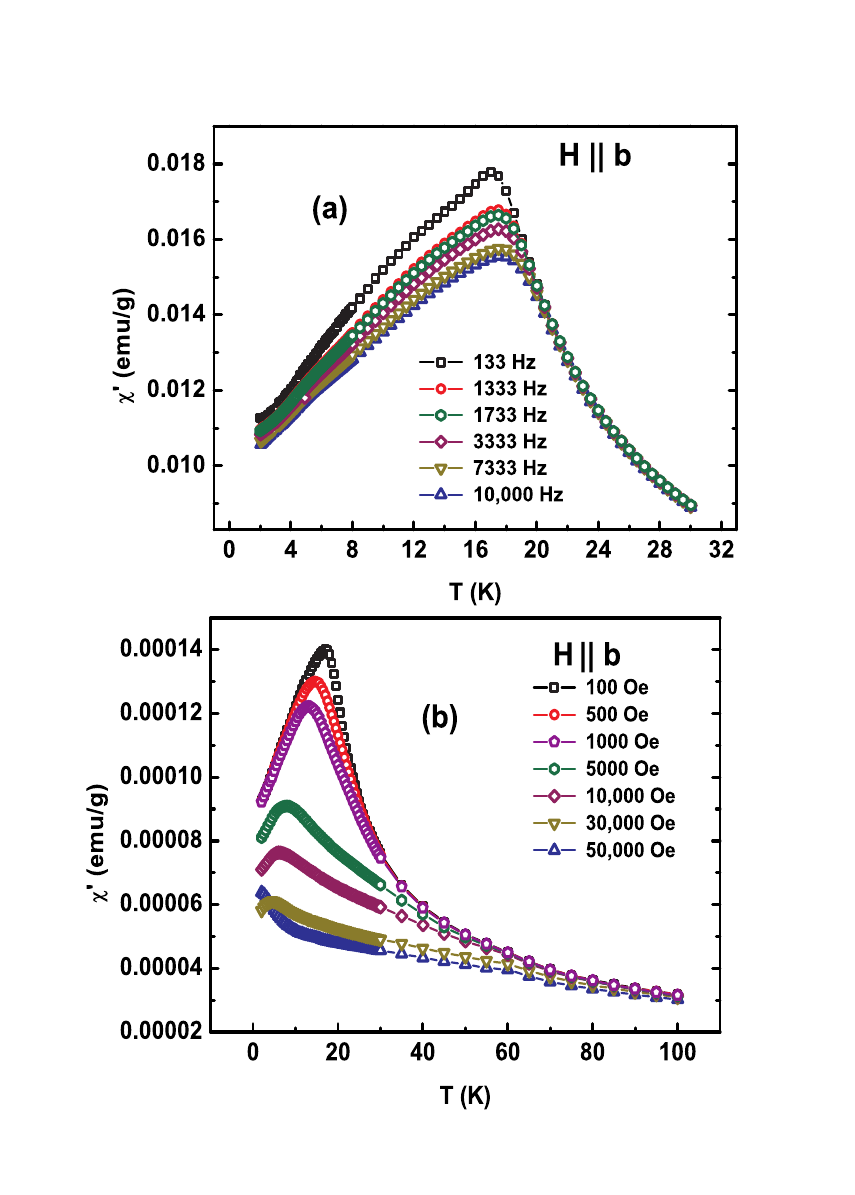}
\caption{(a) The temperature dependence of ac susceptibility $\chi'$ at different frequencies measured with an ac field amplitude of 10 Oe.
(b) The effect of an imposed magnetic field
on the susceptibility of $\textit{o-}$DyMnO$_{3}$ measured in a frequency of 1333 Hz and an ac field amplitude 
of 10 Oe.}
\end{center}
\end{figure}

\clearpage
\begin{figure}
\begin{center}
\includegraphics[scale=1.5]{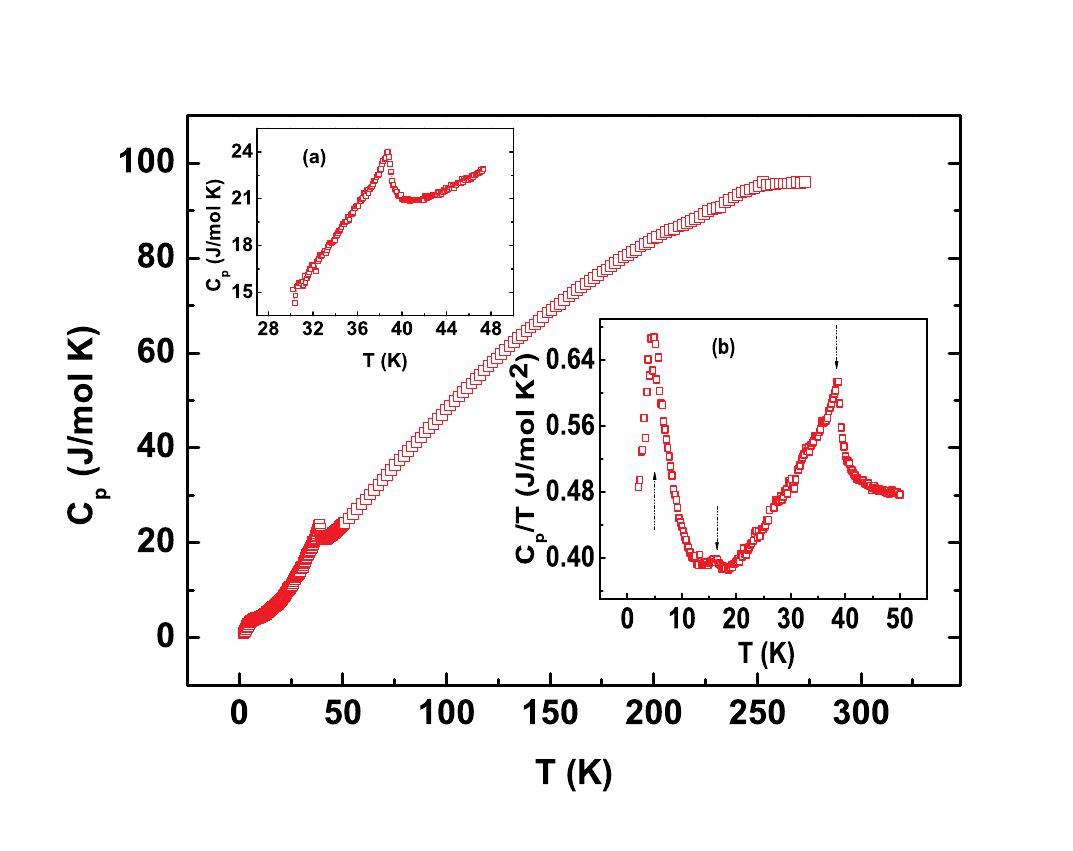}
\caption{Specific heat at zero applied field for
$\textit{o-}$DyMnO$_{3}$. The insets show (a)~the high temperature
peak at the magnetic ordering transition for the Mn-sublattice at
$T_N^{\mathrm Mn}$. (b) the $C_p/T$ plot indicating the
{\textit lock--in} transition at T$_{lock-in} \approx$ 16~K.}
\end{center}
\end{figure}

\end{document}